\documentclass[proceedings]{JHEP} 

\newcommand{\lsim}{\mathrel{\lower4pt\hbox{$\sim$}}
\hskip-12.5pt\raise1.6pt\hbox{$<$}\;}
\newcommand{\gsim}{\mathrel{\lower4pt\hbox{$\sim$}}
\hskip-12.5pt\raise1.6pt\hbox{$>$}\;}

\conference{Third Latin American Symposium on High Energy Physics}

\title{Lessons from Recent Measurements of CP Violation}

\author{Yosef Nir\thanks{Supported by the Israel Science Foundation 
founded by the Israel Academy of Sciences and Humanities, by the 
United States $-$ Israel Binational Science Foundation (BSF) and 
by the Minerva Foundation (Munich).} \\
         Department of Particle Physics, Weizmann
Institute of Science, Rehovot 76100, Israel \\
        E-mail: \email{yosef.nir@weizmann.ac.il}}

\abstract{We discuss various implications of recent experimental results
concerning CP violation and mixing in $K\to\pi\pi$, $B\to\psi K_S$,
$D\to K\pi$ and $D\to KK$ decays.}

\begin{document} 

  \section{Progress in Experiments}
The study of CP violation is entering a new era. For a long period
only a single CP violating observable has been measured \cite{ccft}:
\begin{equation}\label{expeps}
|\varepsilon_K|=2.27\times10^{-3}.
\end{equation}
In the near future, however, we will learn the values of CP asymmetries
in various $B$ decays \cite{casa,bisa,duro}. Then we will be able to test 
the Standard Model (SM) picture of CP violation \cite{koma}. Moreover, 
already at present we know more about CP violation than eq. (\ref{expeps}). 
In particular, in the last year, the following new measurements have been made:

(i) Direct CP violation in $K\to\pi\pi$ decays has been observed. 
The two latest measurements give:
\begin{equation}\label{expepp}
{\varepsilon^\prime\over\varepsilon}=\cases{
(28.0\pm4.1)\times10^{-4}&KTeV \cite{ktev},\cr
(14.0\pm4.3)\times10^{-4}&NA48 \cite{nafe},\cr}
\end{equation}
which, in combination with previous results \cite{NAto,NAton,Esto}, 
give a world average of
\begin{equation}\label{aveepp}
{\varepsilon^\prime\over\varepsilon}=
(1.93\pm0.24)\times10^{-3}.
\end{equation}

(ii) CP violation in $B\to\psi K_S$ has been searched for.
The three most precise measurements give:
\begin{equation}\label{expapk}
a_{\psi K_S}=\cases{
0.79^{+0.41}_{-0.44}&CDF \cite{cdf},\cr
0.12\pm0.37\pm0.09&BaBar \cite{babar},\cr
0.45^{+0.43+0.07}_{-0.44-0.09}&Belle \cite{belle},\cr}
\end{equation}
which give an average of
\begin{equation}\label{aveapk}
a_{\psi K_S}=0.42\pm0.24.
\end{equation}
(For previous measurements, see \cite{opal,cdfo,aleph}.)

(iii) Mixing and CP violation have been searched for in 
doubly Cabibbo suppressed $D\to K\pi$ decays. The
most interesting result concerns the parameter $y^\prime$
which is related to the width difference and the mass difference
between the two neutral $D$ mesons:
\begin{equation}\label{aveypr}
y^\prime=(-2.5^{+1.4}_{-1.6})\times10^{-2}\ \ \ {\rm CLEO}\ \cite{cleo}.
\end{equation}
(For previous, related results, see
\cite{esixno,cleoa,esnosl,esnoa,alephd,esnob}.)
The experimental results for all CP violating observables in 
these decays are consistent with zero.

(iv) Mixing has been searched for by comparing the time
dependence of the $D\to K^+K^-$ decay rates to that of the
Cabibbo favored $D\to K\pi$ decay rates. The parameter $y_{\rm CP}$,
which gives the difference between the respective exponents,
was measured to be 
\begin{equation}\label{aveycp}
y_{\rm CP}=(3.42\pm1.57)\times10^{-2}\ \ \ {\rm FOCUS}\ \cite{focus}.
\end{equation} 

In this review, we will describe various implications of these
new measurements. That should help one to understand how future
measurements will further test the SM and hopefully probe new 
sources of CP violation.

  \section{Open Questions}
\subsection{Features of CP violation}
The Standard Model picture of CP violation is rather unique and
very predictive. Consequently, it can be tested by experiments
in unambiguous ways. Yet, at present CP violation is one of the 
least tested aspects of the Standard Model. The problem lies in 
the fact that there is very little experimental data concerning 
CP violation. 

Here are some of the features of CP violation within the SM:

(i) There is single source of CP violation, that is the 
$\delta_{\rm KM}$-phase. 

(ii) CP violation appears only in the charged current interactions 
of quarks.

(iii) CP violation is closely related to flavor changing interactions.
If the weak interactions were flavor diagonal, the SM would
be CP conserving. 

(iv) CP is not an approximate symmetry of the weak interactions,
$\delta_{\rm KM}={\cal O}(1)$.
The smallness of the two measured parameters, $\varepsilon_K$ 
and ${\varepsilon^\prime}$, is related to the smallness of flavor violation
in the first two generations and not to small phases.

(v) CP is explicitly broken. It arises from complex Yukawa couplings.

(Non-perturbative corrections to the Standard Model tree-level Lagrangian 
are expected to induce $\theta_{\rm QCD}$, a CP violating parameter. This 
second possible source of CP violation is related to strong interactions
and is flavor diagonal. The bounds on the electric dipole moment of the
neutron imply that $\theta_{\rm QCD}\leq10^{-9}$. The Standard Model
offers no natural explanation to the smallness of $\theta_{\rm QCD}$.
We assume that this `strong CP problem' is
solved by some type of new physics, such as a Peccei-Quinn symmetry
\cite{pequ}, which sets $\theta_{\rm QCD}$ to zero.)

It is important to realize that none of features (i)-(v) is experimentally
established and that various reasonable extensions of the SM
provide examples where these features do not hold.
In particular, it could be that CP violation in Nature has 
some or all of the following features:

(i) There are many independent sources of CP violation. For example,
the minimal supersymmetric standard model has forty four independent
phases.

(ii) CP is violated in the lepton sector and/or in neutral
current interactions and/or in new sectors of the theory. In particular,
the recent evidence for neutrino masses \cite{smirnov} makes it very likely 
that CP violation appears in the lepton mixing matrix. Other examples
include supersymmetry \cite{masiero,kanecp,gnr}, where CP violation appears 
also in gluino couplings, that is , in strong interactions. 

(iii) Flavor diagonal interactions violate CP.
For example, in supersymmetry there are new phases in gaugino masses
and in bilinear Higgs couplings that are likely to induce a large
electric dipole moment of the neutron.

(iv) CP is an approximate symmetry. We discuss this possibility in
more detail below.

(v) CP is spontaneously broken.

This situation, where the SM gives a very unique description
of CP violation and experiments have not yet confirmed this description,
is the basis for the strong interest, experimental and theoretical,
in CP violation. There are two types
of unambiguous tests concerning CP violation in the Standard Model:
First, since there is a single source of CP violation, observables
are correlated with each other. For example, the CP asymmetries in
$B\to\psi K_S$ and in $K\to\pi\nu\bar\nu$ are strongly correlated
\cite{bubu,niwo,bepe}.
Second, since CP violation is restricted to flavor changing quark
processes, it is predicted to practically vanish in the lepton sector
and in flavor diagonal processes. For example, the transverse
lepton polarization in semileptonic meson decays, CP violation
in $t\bar t$ production, and (assuming $\theta_{\rm QCD}=0$) the electric 
dipole moment of the neutron are all predicted to be orders of magnitude
below the (present and near future) experimental sensitivity.

To demonstrate how little is the Kobayashi-Maskawa mechanism
of CP violation tested, we now give two explicit examples of classes
of models where CP violation is very different from the SM. 

\subsection{Superweak CP violation}
The original {\it superweak} scenario was proposed in ref. 
\cite{wolfsw}. It stated that CP violation appears in a new
$\Delta S=2$ interaction while there is no CP violation in the SM
$\Delta S=1$ transitions. Consequently, the only observable
CP violating effect is $\varepsilon_K$, while 
$\varepsilon^\prime\sim10^{-8}$ and electric dipole moments (EDMs)
are negligibly small. CP violation via neutral scalar exchange is 
the most commonly studied realization of the superweak idea.

The idea was extended to other meson decays by defining the term 
`superweak CP violation' to imply that there is only indirect CP violation.

{\bf Indirect CP violation} refers to CP violation in
meson decays where the CP violating phases can all be chosen to
appear in $\Delta F=2$ (mixing) amplitudes.

{\bf Direct CP violation} refers to CP violation in
meson decays where some CP violating phases necessarily
appear in $\Delta F=1$ (decay) amplitudes.

If, for example, one extends the superweak scenario to the $B$ system
by assuming that there is CP violation in $\Delta B=2$ but not in $\Delta B=1$
transitions, the prediction for CP asymmetries in $B$ decays into final
CP eigenstates is that they are equal for all final states \cite{winsw,
swsw,wwsw}. In addition, the asymmetries in charged $B$ decays vanish.

As long as $\varepsilon_K$ was the only measured CP violating parameter,
one could assume that CP violation appears only in $\Delta S=2$
transitions, consistent with the superweak scenario. However, the
unambiguous measurements of $\varepsilon^\prime/\varepsilon$
(eq. (\ref{expepp})) have established that there is direct CP violation in
Nature. The superweak idea is consequently excluded. 

\subsection{Approximate CP}
It could be that all CP violating phases are small and that CP is 
an approximate symmetry even of the weak interactions. This idea,
in addition to providing an example for a dramatically different
picture of CP violation compared to the SM, is particularly motivated
in the supersymmetric framework. In a generic supersymmetric
extension of the SM, the supersymmetric contribution to the electric
dipole moment of the neutron $d_N$ is naively expected to be about two
orders of magnitude above the experimental bound. This is the
{\it supersymmetric CP problem}. While the naively large supersymmetric
contributions to $\varepsilon_K$ can be suppressed by flavor-related
mechanisms (such as universality or alignment), this is not the
case for flavor diagonal observables such as $d_N$.
A possible solution of this problem is that all CP violating phases
are small, say $\phi_{\rm CP}\leq{\cal O}(10^{-2})$.

If CP is an approximate symmetry, we expect $\delta_{\rm KM}\ll 1$. 
Then the standard box diagrams cannot account for $\varepsilon_K$ which 
should arise from another source. In supersymmetry with non-universal soft 
terms, the source could be diagrams involving virtual superpartners, 
such as squark-gluino box diagrams. Define $(M_{12}^K)^{\rm SUSY}$ to be
the supersymmetric contribution to the $K-\bar K$ mixing amplitude.
Then the $K-\overline{K}$ mixing constraints give a lower bound
on the size of CP violation: 
\begin{eqnarray}\label{lowapp}
{\cal R}e (M_{12}^K)^{\rm SUSY}&\lsim&\Delta m_K,\nonumber\\
{\cal I}m (M_{12}^K)^{\rm SUSY}&\sim&\varepsilon_K\Delta m_K\nonumber\\
\Longrightarrow \phi_{\rm CP}\geq{\cal O}(\varepsilon_K)&\sim& 
10^{-3}.
\end{eqnarray} 
As mentioned above, the $d_N$ constraint \cite{pdg} gives an upper bound
on the size of CP violation:
\begin{equation}\label{uppapp}
d_N\leq6.3\times10^{-26}\ e\ {\rm cm}^{-2}\ \Longrightarrow\  \phi_{\rm CP}
\lsim10^{-2}.
\end{equation} 
If all phases are of the same order, then $d_N$ must be
just below or barely compatible with the present experimental bound.
A signal should definitely be found if the accuracy is increased by two
orders of magnitude.
 
The main phenomenological implication of these scenarios is that
CP asymmetries in $B$ meson decays are small, perhaps
${\cal O}(\varepsilon_K)$, rather than ${\cal O}(1)$ as expected in the SM. 
Explicit models of approximate CP were presented in refs. \cite{poma,dkl,baba,
abfr,eyni,bdm}. Some of these models cannot accommodate the value of 
$\varepsilon^\prime/\varepsilon$. We will return to this question later.  

The situation that both the SM and models of approximate
CP are viable at present is related to the fact that the
mechanism of CP violation has not really been tested experimentally.
The only measured CP violating observables, $\varepsilon_K$ and
$\varepsilon^\prime/\varepsilon$, are small. Their
smallness can be related to the accidental smallness of CP violation
for the first two quark generations, as in the Standard
Model, or to CP being an approximate symmetry, as in the
models discussed here. Future measurements, particularly of processes
where all three generations play a role (such as $B\to\psi K_S$
or $K\to\pi\nu\bar\nu$), will easily distinguish between the two scenarios.
While the Standard Model predicts large CP violating effects
for these processes, approximate CP would suppress them too.

  \section{Lessons from $\varepsilon^\prime/\varepsilon$}
\subsection{$\varepsilon^\prime/\varepsilon$ in supersymmetric models}
The $\varepsilon^\prime$ parameter, signifying direct CP violation, 
has now been measured with impressive accuracy. 
The theoretical interpretation of this result suffers from large
hadronic uncertainties. Within the Standard Model, the theoretically preferred
range is somewhat lower than the experimental range of eq. (\ref{aveepp})
(for recent reviews, see \cite{bef,buraep,nierep,jamin,cima}). Yet, if all 
the hadronic parameters are taking values at the extreme of their reasonable 
ranges, the experimental result can be accommodated. 

While (\ref{aveepp}) does not provide unambiguous evidence for new physics, 
it is still useful in testing extensions of the Standard Model.
Models where $\varepsilon^\prime$ is suppressed and/or $\varepsilon_K$ 
enhanced are disfavored. Models that allow significant new contributions to
$\varepsilon^\prime$ may be favored if future improvements in the 
theoretical calculation will 
prove that the Standard Model fails to account for its large value.
Investigations of the supersymmetric contributions to 
$\varepsilon^\prime/\varepsilon$ in view of the recent experimental results 
have been presented in refs. \cite{MaMu,KhKo,BJKP,BCS,emns,kane,bekkl,dmv,kkv,
bggjss}. Here we review the work of ref. \cite{emns} which focuses on
flavor models with approximate CP. In models where CP is an approximate 
symmetry of electroweak interactions, that is, all CP violating phases are 
small, it is clear already at present that the Standard Model cannot 
explain (\ref{aveepp}). These models should then provide new contributions to
fully account for $\varepsilon^\prime/\varepsilon$. A failure to do so would 
mean that the model is excluded. We review here one such class of models,
analyzed in ref. \cite{emns}, where
a horizontal Abelian symmetry solves the supersymmetric flavor problems by means
of alignment and approximate CP solves the remaining CP problems.

In generic supersymmetric models, there are potentially many new contributions 
to $\varepsilon^\prime/\varepsilon$ from loop diagrams involving
intermediate squarks and gluinos, charginos or neutralinos.
If there is some degeneracy between squarks, then a convenient way 
to parameterize these contributions is by using the $(\delta_{MN}^q)_{ij}$ 
parameters. In the basis where quark masses and gluino couplings are diagonal, 
the dimensionless $(\delta_{MN}^q)_{ij}$ parameters
stand for the ratio between $(M^2_{\tilde q})^{MN}_{ij}$, the $(ij)$ 
entry ($i,j=1,2,3$) in the mass-squared matrix for squarks ($M,N=L,R$ 
and $q=u,d$), and $\tilde m^2$, the average squark mass-squared.
If there is no mass degeneracy among squarks, then these parameters
can be related to the supersymmetric mixing angles. Defining $K_L^d$ 
($K_R^d$) to be the mixing matrix between left (right) handed down 
quarks and the scalar partners of left (right) handed down quarks,
we have, {\it e.g.}, $(\delta_{LL}^d)_{12}\sim(K_L^d)_{12}$. 

For supersymmetry to account for $\varepsilon^\prime/\varepsilon$, at least one 
of the following six conditions should be met \cite{GMS,GGMS,CoIs,BuSi}:
\begin{eqnarray}\label{susyepe}
{\cal I}m[(\delta^d_{LL})_{12}]&\sim&
\lambda\left({\tilde m\over500\ GeV}\right)^2,\nonumber\\
{\cal I}m[(\delta^d_{LR})_{12}]&\sim&
\lambda^7\left({\tilde m\over500\ GeV}\right),\nonumber\\
{\cal I}m[(\delta^d_{LR})_{21}]&\sim&\lambda^7
\left({\tilde m\over500\ GeV}\right),
\end{eqnarray}
\begin{eqnarray}\label{Zdsepe}
{\cal I}m[(\delta^u_{LR})_{13}(\delta^u_{LR})_{23}^*]&\sim&
\lambda^2,\nonumber\\
{\cal I}m[V_{td}(\delta^u_{LR})_{23}^*]&\sim&
\lambda^3\left({M_2\over m_W}\right),\nonumber\\
{\cal I}m[V_{ts}^*(\delta^u_{LR})_{13}]&\sim&
\lambda^3\left({M_2\over m_W}\right).
\end{eqnarray}
Here $\lambda=0.2$ is a small parameter of order of the Cabibbo angle
that is convenient to use in the context of flavor models.

Let us first discuss the three options of eq. (\ref{susyepe}). The first of 
these conditions violates constraints from $\Delta m_K$ and $\varepsilon_K$. 
Therefore, independent of the supersymmetric flavor model,
it cannot be satisfied. On the other hand, the requirements on
${\cal I}m[(\delta^d_{LR})_{12}]$ or ${\cal I}m[(\delta^d_{LR})_{21}]$ pose
no phenomenological problem. Moreover, we will see in the next subsection
that such values are possible within our theoretical framework. 
We therefore investigate more carefully the uncertainties in the 
corresponding condition. Using the expression for the matrix element of the
chromomagnetic operator from \cite{BEF} and defining a parameter $B_G$ 
to account for possible deviations from its value obtained at lowest
order in the chiral quark model, one can write \cite{BCIRS}:  
\begin{eqnarray}\label{LRdot}
&\left|{\varepsilon^\prime\over\varepsilon}\right|&=58\ B_G\left[
{\alpha_s(m_{\tilde g})\over\alpha_s(500\ GeV)}\right]^{23/21}\left(
{158\ MeV\over m_s+m_d}\right)\nonumber\\ &\times&\left(
{500\ GeV\over m_{\tilde g}}\right)\left|{\cal I}m
\left[(\delta^d_{LR})_{12}-(\delta^d_{LR})^*_{21}\right]\right|.
\end{eqnarray}
Using rather conservative estimates \cite{emns}, one gets a lower bound:
\begin{equation}\label{lowerLRot}
{\cal I}m(\delta^d_{LR})_{12}\gsim7\times10^{-7},
\end{equation}
that is ${\cal O}(\lambda^9)$ or even ${\cal O}(\lambda^{10})$ if 
$\lambda\sim0.24$. A similar bound applies to ${\cal I}m[(\delta^d_{LR})_{21}]$.

We now turn to the three options in eq. (\ref{Zdsepe}). These contributions
to $\varepsilon^\prime/\varepsilon$ arise by inducing an effective $Z_{ds}$ 
coupling, where \cite{BuSi}:
\begin{equation}\label{defZds}
{\cal L}_{\rm FC}^Z={G_F\over\sqrt{2}}{em_Z^2\over2\pi^2}
{\cos\theta_W\over\sin\theta_W}Z_{ds}\bar s\gamma_\mu(1-\gamma_5)dZ^\mu
+{\rm h.c.}.
\end{equation}
The contribution of such an effective coupling to 
$\varepsilon^\prime/\varepsilon$ is given by
\begin{equation}\label{epeZds}
{\varepsilon^\prime\over\varepsilon}={\cal I}m Z_{ds}\left[1.2-\left(
{158\ MeV\over m_s+m_d}\right)^2\left|r_Z^{(8)}\right|B_8^{(3/2)}\right],
\end{equation}
where $B_8^{(3/2)}$ is the non-perturbative parameter describing
the hadronic matrix element of the electroweak penguin operator and
$\left|r_Z^{(8)}\right|$ is a calculable renormalization scheme
independent parameter. Using the ranges given in \cite{BuSi}, one gets:
\begin{equation}\label{epeonZds}
-{\cal I}m Z_{ds}\gsim(\varepsilon^\prime/\varepsilon)/16,
\end{equation}
leading to
\begin{eqnarray}\label{weakZds}
{\cal I}m[V_{td}(\delta^u_{LR})_{23}^*]&\gsim&2\times10^{-3}\sim\lambda^4,
\nonumber\\
{\cal I}m[V_{ts}(\delta^u_{LR})_{13}^*]&\gsim&2\times10^{-3}\sim\lambda^4,
\nonumber\\
{\cal I}m[(\delta^u_{LR})_{13}(\delta^u_{LR})_{23}^*]&\gsim&
\lambda^3.
\end{eqnarray}

\subsection{Abelian flavor symmetries}
Models of Abelian horizontal symmetries are able to provide a
natural explanation for the hierarchy in the quark and lepton
flavor parameters \cite{FrNi}.
The symmetry is broken by a small parameter $\lambda$ which is usually
taken to be of the order of the Cabibbo angle, $\lambda\sim0.2$.
The hierarchy in the flavor parameters is then a result of the
selection rules related to the approximate horizontal symmetry.
In the supersymmetric framework, holomorphy also plays a role in
determining the Yukawa parameters \cite{LNS}.

A typical structure of the quark mass matrices in such model is as follows:
\begin{eqnarray}\label{qmama}
M_u&\sim&\langle\phi_u\rangle\pmatrix{\lambda^7&\lambda^5&\lambda^3\cr
\lambda^6&\lambda^4&\lambda^2\cr \lambda^4&\lambda^2&1\cr},\nonumber\\ 
M_d&\sim&\langle{\phi_d}\rangle\lambda^3\tan\beta
\pmatrix{\lambda^4&\lambda^3&\lambda^3\cr
\lambda^3&\lambda^2&\lambda^2\cr \lambda&1&1\cr}.
\end{eqnarray}
A similar hierarchy appears also in the $(LR)$ blocks of the corresponding
squark mass-squared matrices:
\begin{equation}\label{sqLR}
(M^2_{\tilde u})^{LR}_{ij}\sim \tilde m(M_u)_{ij},\ \ \ 
(M^2_{\tilde d})^{LR}_{ij}\sim \tilde m(M_d)_{ij}.
\end{equation}

Eqs. (\ref{qmama}) and (\ref{sqLR}) allow us to estimate the values of the 
$\delta_{LR}$ parameters of eq. (\ref{susyepe}) and (\ref{Zdsepe}). We get:
\begin{eqnarray}\label{houoepe}
(\delta^d_{LR})_{12}&\sim&{m_s|V_{us}|\over\tilde m}\sim\lambda^6\
{m_t\over\tilde m},\nonumber\\ 
(\delta^d_{LR})_{21}&\sim&{m_d\over|V_{us}|\ \tilde m}\sim\lambda^6\
{m_t\over\tilde m},\nonumber\\
(\delta^u_{LR})_{13}&\sim&{m_t|V_{ub}|\over\tilde m}\sim\lambda^3\
{m_t\over\tilde m},\nonumber\\
(\delta^u_{LR})_{23}&\sim&{m_t|V_{cb}|\over\tilde m}\sim\lambda^2\
{m_t\over\tilde m}.
\end{eqnarray}

Taking into account that $|V_{td}|\sim\lambda^3$ and $|V_{ts}|\sim\lambda^2$,
we learn that the three options in eq. (\ref{Zdsepe}) are of order
$\lambda^{5-7}$. We compare this to the requirements given in eq. 
(\ref{weakZds}) and conclude that, in models
of Abelian horizontal symmetries, the contributions to $Z_{ds}$ involving
$\tilde t_R$ cannot account for $\varepsilon^\prime/\varepsilon$.

On the other hand, $(\delta^d_{LR})_{12}$ and $(\delta^d_{LR})_{21}$ are 
large enough to allow for a supersymmetric explanation of 
$\varepsilon^\prime/\varepsilon$ \cite{MaMu}.

\subsection{Alignment and approximate CP}
It is possible to solve the supersymmetric flavor problems by the
mechanism of alignment \cite{NiSe,LNSb,GrNiL,RaNi}, whereby the mixing 
matrices for gaugino couplings have very small mixing angles. 
Alignment arises naturally in the framework of Abelian horizontal symmetries. 
Simple models give supersymmetric mixing angles that are similar to the 
corresponding CKM mixing angles. However, for the mixing between the first two
down squark generations, a much more precise alignment is 
phenomenologically needed:
\begin{eqnarray}\label{Kdot}
(K_L^d)_{12}\lsim\lambda^2,\ \ \ (K_R^d)_{12}&\lsim&\lambda^2,\nonumber\\
(K_L^d)_{12}(K_R^d)_{12}&\lsim&\lambda^6.
\end{eqnarray}
To achieve the required suppression, one has to employ a more complicated
Abelian horizontal symmetry. The models of refs. \cite{NiSe,LNSb,GrNiL,RaNi}
use $U(1)\times U(1)$ symmetries. Then, it is possible to retain all the
`good' predictions for the quark mass ratios and the CKM mixing angles and, 
at the same time, have Yukawa couplings that are relevant to (\ref{Kdot}) 
vanish due to holomorphy of the superpotential, that is, $(M_d)_{12}=0$, 
$(M_d)_{21}=0$, either $(M_d)_{13}$ or $(M_d)_{32}=0$ and either $(M_d)_{31}$ 
or $(M_d)_{23}=0$. 

The entries in the $LR$-block of the down-squark mass-squared matrix,
that is $(M^2_{\tilde d})^{LR}_{ij}$, are suppressed in a similar way to
the corresponding entries in the down quark mass matrix, $(M_d)_{ij}$.
Consequently, the alignment requirements (\ref{Kdot}) affect directly 
$(M^2_{\tilde d})^{LR}_{12}$ and $(M^2_{\tilde d})^{LR}_{21}$ that
are relevant to $\varepsilon^\prime/\varepsilon$. Independent of the details 
of the model, we find that in the framework of alignment, we have
\begin{eqnarray}\label{aligepe}
(\delta^d_{LR})_{12}&\lsim&{m_s|V_{us}|\over\tilde m}\lambda^2\sim\lambda^8\
{m_t\over\tilde m},\nonumber\\
(\delta^d_{LR})_{21}&\lsim&{m_d\over|V_{us}|\ \tilde m}\lambda^2\sim\lambda^8\
{m_t\over\tilde m}.
\end{eqnarray}
The values in eq. (\ref{aligepe}) should be compared with the phenomenological 
input of eq. (\ref{susyepe}). It is interesting that for central values of the 
hadronic parameters, the supersymmetric contributions to 
$\varepsilon^\prime/\varepsilon$ in models of
alignment can naturally be of the required order of magnitude.
For this to happen, the models have to satisfy two conditions:
\begin{itemize}
\item The alignment has to be minimal,
that is either $|(K_L^d)_{12}|\sim\lambda^3$ or 
$|(K_R^d)_{12}|\sim\lambda^3$ should hold.
\item The relevant phase is of order one.
\end{itemize}

We now focus on models where all flavor problems are solved by
alignment, but the CP problems are solved by approximate CP.
(In a different class of models, the alignment is precise enough 
to solve also some of the CP problems \cite{RaNi}.)
The main point is that, independent of the details of the model,
the CP violating phases in this framework are suppressed by even
powers of the breaking parameter. Consequently, the imaginary
part of any $(\delta^q_{MN})_{ij}$ term is suppressed by, at least,
a factor of $\lambda^2$ compared to the real part. In particular, we have
\begin{eqnarray}\label{apprepe}
{\cal I}m(\delta^d_{LR})_{12}&\lsim&{m_s|V_{us}|\over\tilde m}\lambda^4
\sim\lambda^{10}{m_t\over\tilde m},\nonumber\\
{\cal I}m(\delta^d_{LR})_{21}&\lsim&{m_d\over|V_{us}|\ \tilde m}\lambda^4\sim
\lambda^{10}{m_t\over\tilde m}.
\end{eqnarray}
These are rather low values. They are consistent with the experimental
constraint of eq. (\ref{lowerLRot}) only if all the following conditions are
simultaneously satisfied:
\begin{itemize}
\item The suppression of the relevant CP violating phases is `minimal', 
$\phi_{\rm CP}={\cal O}(\lambda^2)$.
\item The alignment of the first two down squark generations is 
`minimal', $|(K^d_M)_{12}|={\cal O}(\lambda^3)$ where $M=L$ or $R$.
\item The mass scale for the supersymmetric particles is low, 
$\tilde m\sim150\ GeV$.
\item The hadronic matrix element is larger than what hadronic
models suggest, $B_G\sim5$.
\item The mass of the strange quark is at the lower side of the
theoretically preferred range, $m_s(m_c)\sim110\ MeV$.
\item The value of $\varepsilon^\prime/\varepsilon$ is at the lower side 
of the experimentally allowed range.
\end{itemize}

While such a combination of conditions on both the supersymmetric
models and the hadronic parameters is not very likely to be realized,
it cannot be rigorously excluded either. We conclude that models
that combine alignment and approximate CP are disfavored by the
measurement of $\varepsilon^\prime/\varepsilon$.

  \section{Lessons from $a_{\psi K_S}$}
\subsection{Introduction}
Experiments are closing in on the value of the CP asymmetry
in $B\to\psi K_S$ \cite{cdf,babar,belle}. The CP violating quantity 
$a_{\psi K_S}$ is defined through
\begin{eqnarray}\label{defapks}
A_{\psi K_S}(t)&\equiv&
{\Gamma[\overline{B^0}(t)\to\psi K_S]-\Gamma[B^0(t)\to\psi K_S]
\over \Gamma[\overline{B^0}(t)\to\psi K_S]+\Gamma[B^0(t)\to\psi K_S]}
\nonumber\\ &=&
a_{\psi K_S}\sin(\Delta m_B t).
\end{eqnarray}
Within the Standard Model, $a_{\psi K_S}$ is related to the angle $\beta$
of the unitarity triangle,
\begin{equation}\label{defbet}
a_{\psi K_S}=\sin2\beta,\ \ \ \beta\equiv\arg\left[-
{V_{cd}V_{cb}^*\over V_{td}V_{tb}^*}\right].
\end{equation}
Based on the determination of the CKM parameters through measurements
of $|V_{ub}/V_{cb}|$, $\varepsilon_K$, $\Delta m_B$ and $\Delta m_{B_s}$,
the Standard Model prediction is
\begin{equation}\label{numstb}
0.59\lsim\sin2\beta\lsim0.82.
\end{equation}
Thus, the measurement of $a_{\psi K_S}$ (\ref{aveapk}) is consistent with the 
SM prediction (\ref{numstb}). Yet, the allowed range in (\ref{aveapk}) leaves
open the possibility that $a_{\psi K_S}$ is actually significantly smaller
than the SM prediction. This possibility was recently investigated in refs. 
\cite{kaneb,siwob,enp,xing}. Here we review the work of ref. \cite{enp}.
For the sake of concreteness, it was assumed in ref. \cite{enp} that 
$a_{\psi K_S}$  lies below the 1$\sigma$ upper bound of the BaBar measurement 
\cite{babar},
\begin{equation}\label{numapks}
a_{\psi K_S}\lsim0.5.
\end{equation}

If, indeed, $a_{\psi K_S}\leq0.5$, there are two ways in which the
conflict with (\ref{numstb}) might be resolved:
\begin{itemize}
\item The SM is valid but one or more of the hadronic parameters 
which play a role in the analysis that leads to (\ref{numstb}) are outside
their `reasonable range'. 
\item New physics affects the CP asymmetry in $B\to\psi K_S$
and/or some of the measurements that lead to (\ref{numstb}).
\end{itemize}
Below we discuss these two possibilities. 

\subsection{Hadronic uncertainties}
The computations that relate experimental observables to
CKM parameters suffer, in general, from theoretical uncertainties
\cite{quinn}.
In very few cases, the calculation is made entirely in the
framework of a systematic expansion and it is possible to
reliably estimate the error that is induced by truncating the
expansion at a finite order. This is the case with the
relation between the observable $a_{\psi K_S}$ and the CKM parameter
$\sin2\beta$: Within the SM, the relation (\ref{defbet}) holds to
an accuracy of better than one percent (for a review, see \cite{nirssi}).
Thus, if we assume that (\ref{numapks}) holds, we have
\begin{equation}\label{acpcon}
{2\bar\eta(1-\bar\rho)\over\bar\eta^2+(1-\bar\rho)^2}\leq+0.50.
\end{equation}

In most cases, however, the calculation involves models or, 
on occasion, educated guesses and there is no
easy way to estimate the errors that are involved. This is
the case with almost all the observables that are involved
in the prediction (\ref{numstb}). We follow the treatment of
this issue of ref. \cite{babook}. We will quote `reasonable ranges'
for the parameters that involve uncontrolled theoretical
uncertainties, and compare them to the values that are 
required for consistency with (\ref{numapks}).

As mentioned above, the prediction in eq. (\ref{numstb}) is
based on four observables:

{\bf (i) Charmless semileptonic $B$ decays} determine the 
$R_u$ parameter:
\begin{equation}\label{ruexp}
R_u\equiv\sqrt{\bar\rho^2+\bar\eta^2}
={1\over\lambda}\left|{V_{ub}\over V_{cb}}\right|.
\end{equation}
There is a large uncertainty  in this determination, coming from 
hadronic modeling of the decays. On the other hand, the fact that 
one gets consistent results from both inclusive and various exclusive
measurements makes it reasonable to think that the error on
$|V_{ub}/V_{cb}|^2$ is not larger than order 40\%. A reasonable
range is then $R_u=0.39\pm0.07$. If, however, the 
inconsistency  between (\ref{numapks}) and (\ref{numstb}) comes
entirely from the hadronic modeling of charmless semileptonic $B$ decays, 
the failure of these models should be such that $|V_{ub}/V_{cb}|$ is about 
30\% lower than the presently most favorable value, $R_u\lsim0.27$.

{\bf (ii) The mass difference between the two neutral $B$ mesons},
$\Delta m_B$, determines the $R_t$ parameter: 
\begin{equation}\label{rtexp}
R_t\equiv\sqrt{(1-\bar\rho)^2+\bar\eta^2}
={1\over\lambda}\left|{V_{td}\over V_{ts}}\right|.
\end{equation}
This determination suffers from a large hadronic uncertainty
in the matrix element of the relevant four quark operator
which is parameterized by $\sqrt{B_{B_d}}f_{B_d}$.

The ratio $\Delta m_{B_s}/\Delta m_{B_d}$ can also be
used to determine $R_t$. Since at present there is only a lower
bound on $\Delta m_{B_s}$, this ratio provides only an upper bound
on $R_t$. This upper bound is stronger however than the one derived
from $\Delta m_{B_d}$ alone. Moreover, it suffers from smaller
hadronic uncertainties since the ratio 
\begin{equation}\label{defxi}
\xi\equiv{\sqrt{B_{B_s}}f_{B_s}\over\sqrt{B_{B_d}}f_{B_d}}
\end{equation}
is one in the SU(3) limit. Lattice calculations have only to determine
the SU(3) breaking effect, that is the deviation from $\xi=1$.
This calculation is believed to be under better control and
an uncertainty of order 50\% on the deviation from one seems reasonable. 
The one sigma range is then $\xi=1.14\pm0.08$. If, however, the 
inconsistency  between (\ref{numapks}) and (\ref{numstb}) comes
entirely from an error in the estimate of $\xi$, the failure of 
lattice calculations should be such that $\xi-1$ is at least  
a factor of three larger than the presently most favorable value, 
$\xi\gsim1.4$.
 
{\bf (iii) CP violation in neutral kaon mixing,
$\varepsilon_K$} gives another constraint in the $(\bar\rho,\bar\eta)$
plane. The main source of uncertainty is in $\hat B_K$ which
parameterizes the matrix element of the four quark operator.
While each method for its determination suffers from uncontrolled
theoretical errors, the fact that many different methods give
similar ranges makes it reasonable to assign to it a theoretical
error of order 15-20\%. The reasonable range is then $\hat B_K=0.80\pm0.15$.
If, however, the inconsistency with $a_{\psi K_S}$ comes
entirely from the $\varepsilon_K$ constraint, the failure of 
the various calculations should be such that $\hat B_K$ is at least  
50\% larger than the presently most favorable value, $\hat B_K\gsim1.3$.

To summarize: assuming that the CP asymmetry in $B\to\psi K_S$
is below the $1\sigma$ upper bound of BaBar measurement,
the SM could still be valid if some of the hadronic parameters
are outside of their `reasonable' ranges. If the apparent discrepancy 
is related to an error in the theoretical estimate of just one parameter,
then it requires either a small value of $|V_{ub}|$, or a large value
of $\xi$ or a large value of $\hat B_K$. The first of these,
$|V_{ub}/V_{cb}|\lsim0.06$, is perhaps the least unlikely deviation
from our `reasonable ranges.'

\subsection{New physics}
New physics can explain an inconsistency of $a_{\psi K_S}$ 
measurement with the SM predictions. It can do so provided
that it contributes significantly either to $B-\overline{B}$
mixing or to the CP violating part of $K-\overline{K}$ mixing
or to both. In this section we examine each of these possibilities. 

It is also possible, in principle, that the discrepancy is explained 
by a new contribution to $b\to u\ell\nu$ decays or to 
$b\to c\bar cs$ decays. We find it unlikely, however, that these SM 
tree level decays are significantly affected by new physics. 

{\bf (i) $B^0-\overline{B^0}$ mixing:}

The effects of new physics on $B^0-\overline{B^0}$ mixing can be 
parameterized as follows \cite{SoWo,DDO,SiWo,CKLN,GNW}:
\begin{equation}\label{motnp}
M_{12}=r_d^2 e^{2i\theta_d}M_{12}^{\rm SM}.
\end{equation}
Here $M_{12}$ ($M_{12}^{\rm SM}$) is the full (SM) 
$B^0-\overline{B^0}$ mixing amplitude. 

If the new physics modifies the phase of the mixing amplitude,
$2\theta_d\neq0$, then the CP asymmetry in $B\to\psi K_S$ is 
modified. Instead of eq. (\ref{defbet}) we now have:
\begin{equation}\label{apksnp}
a_{\psi K_S}=\sin2(\beta+\theta_d).
\end{equation}
If the new physics modifies the magnitude of the mixing 
amplitude, $r_d^2\neq1$, then $\Delta m_B$ is modified,
$\Delta m_B=r_d^2\Delta m_B^{\rm SM}$. In addition, if the new physics 
modifies the $B_s-\overline{B_s}$ mixing amplitude, and we parameterize 
this modification with corresponding parameters $r_s^2$ and $2\theta_s$, 
then 
\begin{equation}\label{ratdsnp}
\Delta m_{B_s}/\Delta m_{B_d}=(r_s/r_d)^2
(\Delta m_{B_s}/\Delta m_{B_d})^{\rm SM}.
\end{equation}

If there is no new physics in the $R_u$ and $\varepsilon_K$ 
constraints, then to achieve consistency with the measurements concerning 
$B-\overline{B}$ mixing, it is required that either (i) $r_d/r_s\neq1$,
or (ii) $2\theta_d\neq0$ or (iii) both. In particular,
consider models where there is a new contribution to both 
$B^0-\overline{B^0}$ mixing and $B_s-\overline{B_s}$ mixing, 
but at least the first of these carries the same phase as the 
Standard Model contribution. With $2\theta_d=0$, we need 
\begin{equation}\label{rnpb}
0.5\lsim r_d\lsim1,\ \ \ r_s/r_d\gsim1.1.
\end{equation}
We conclude that if the new physics contribution carries no new phase, then
it must be flavor violating in the sense that $r_s\neq r_d$. In other words, 
the flavor structure should be different from the CKM one.

Another interesting point is that, to accommodate (\ref{numapks}), 
the contribution of the new physics to the mixing amplitude cannot be much 
smaller than the Standard Model one, 
\begin{equation}\label{npvssm}
|M_{12}^{\rm NP}/M_{12}^{\rm SM}|\gsim0.1.
\end{equation}

{\bf (ii) $K^0-\overline{K^0}$ mixing:}

A value of $a_{\psi K_S}$ below the SM prediction can arise
even if there is no new physics in $B-\overline{B}$ mixing and in
$b\to c\bar cs$ decay, the two processes that are relevant 
to the CP asymmetry in $B\to\psi K_S$.
The explanation must then be related to processes (other
than $\Delta m_B$) that play a role in constraining the 
$\sin2\beta$ range. The prime suspect is CP violation in the
neutral kaon system, that is $\varepsilon_K$.

With no significant new physics contributions to the mixing and the 
relevant decays of the $B$-mesons, the $R_u$, $\Delta m_{B_q}$ and 
$a_{\psi K_S}$ constraints hold. In such a case
there is a small region around $(\bar\rho,\bar\eta)=(0.25,0.20)$
that is marginally consistent with all of these constraints when 
the hadronic parameters reside within our `reasonable ranges.'
In this region, the new physics has to add up constructively
to the SM contribution to $\varepsilon_K$, with 
\begin{equation}\label{npeps}
{\cal I}m\ M_{12}^{\rm NP}(K)/{\cal I}m\ M_{12}^{\rm SM}(K)\gsim0.3.
\end{equation}
 
The situation where the $R_u$, $\Delta m_{B_s}/\Delta m_{B_d}$ and 
$a_{\psi K_S}$ constraints are valid but the $\varepsilon_K$ and 
$\Delta m_{B_d}$ constraints are not
arises in models of new physics where all flavor violation 
and CP violation are described by the CKM matrix. This class of models
was defined and analyzed in ref. \cite{bggjs}. Ref. \cite{enp} finds that, 
if (\ref{numapks}) holds, new physics should play a role in 
both $\varepsilon_K$ and $\varepsilon^\prime/\varepsilon$.

{\bf (iii) Neutral meson mixing:}

In a large class of models, there could be significant contributions
to both $B-\overline{B}$ mixing and $K-\overline{K}$ mixing. However, 
$b\to u\ell\nu$ decays are dominated by the $W$-mediated tree level decay.
The implications of measurements of $a_{\psi K_S}$ in such a framework
were recently investigated in refs. \cite{ben,EyNiapk}.
In such a framework, only the $R_u$ constraint holds leading to
\begin{equation}\label{sintb}
|\sin2\beta|\leq0.82.
\end{equation}
There is a large range of $r_d^2$ and $2\theta_d$ that can accommodate
a low $a_{\psi K_S}$.

Further complications in the analysis occur if there are extra quarks beyond 
the three generations of the SM. In such a case, there are more ways in which 
the CKM constraints can be modified \cite{EyNiapk}. However,  
the dominant effect is always a new contribution to the mixing \cite{NiSi}.

Finally, we note that the presently allowed range for $a_{\psi K_S}$
is consistent with zero asymmetry at the 1.75$\sigma$ level and certainly 
does not exclude the possibility that the asymmetry is small. This leaves 
viable the framework discussed in section 2.3 where CP is an approximate 
symmetry of the full theory, that is, CP violating phases are all small. 
 
  \section{Lessons from $D-\overline{D}$ Mixing Parameters}
\subsection{Formalism}
Recent studies of time-dependent decay rates of $D^0\rightarrow K^+\pi^-$ 
by the CLEO collaboration \cite{cleo} and measurements of the combination of 
$D^0\rightarrow K^+K^-$ and $D^0\rightarrow K^-\pi^+$ rates by the FOCUS
collaboration \cite{focus} have provided highly interesting results
concerning $D^0-\overline{D^0}$ mixing. Each of the two experiments
finds a signal for mixing at a level that is close to $2\sigma$. 
It is not unlikely that these signals are just the results of statistical 
fluctuations and the true mixing parameters lie well below the experimental
sensitivity. It is interesting however to analyze the implications of
the experimental results assuming that their central values are not far 
from the true values and that $D-\overline{D}$ mixing has indeed been 
observed. Such a task has been taken in ref. \cite{bglnp} which we
review here.

We investigate neutral $D$ decays. The two mass eigenstates,
$|D_1\rangle$ of mass $m_1$ and width $\Gamma_1$ and
$|D_2\rangle$ of mass $m_2$ and width $\Gamma_2$,
are linear combinations of the interaction eigenstates:
\begin{eqnarray}\label{masint}
|D_1\rangle\ &=&\ p|D^0\rangle+q|\overline{D^0}\rangle,\nonumber\\
|D_2\rangle\ &=&\ p|D^0\rangle-q|\overline{D^0}\rangle.
\end{eqnarray}
The average mass and width are given by
\begin{equation}\label{SumMG}
m\equiv {m_1+m_2\over2},\ \ \ 
\Gamma\equiv{\Gamma_1+\Gamma_2\over2}.
\end{equation}
The mass and width difference are parameterized by
\begin{equation}\label{DelMG}
x\equiv{m_2-m_1\over\Gamma},\ \ \ 
y\equiv{\Gamma_2-\Gamma_1\over2\Gamma}.
\end{equation}
Decay amplitudes into a final state $f$ are defined by
\begin{equation}\label{AbarA}
A_f\equiv\langle f|{\cal H}_d|D^0\rangle,\ \ \ 
\bar A_f\equiv\langle f|{\cal H}_d|\overline{D^0}\rangle.
\end{equation}
It is useful to define the complex parameter $\lambda_f$:
\begin{equation}\label{deflam}
\lambda_f\equiv{q\over p}\ {\bar A_f\over A_f}.
\end{equation}

Within the Standard Model, the physics of $D-\overline{D}$ mixing and of the
tree level decays is dominated by the first two generations and,
consequently, CP violation can be safely neglected. In all `reasonable'
extensions of the Standard Model, the relevant tree level $D$ decays  
are still dominated by the Standard Model CP 
conserving contributions \cite{BSN,BeNi}. On the other hand, there could be 
new short distance, possibly CP violating contributions to the mixing amplitude 
$M_{12}$. Allowing for only such effects of new physics, the picture of CP 
violation is simplified since there is no direct CP violation. The effects of 
indirect CP violation can be parameterized in the following way \cite{nirssi}:
\begin{eqnarray}\label{parlkpi}
|q/p|&=&R_m,\nonumber\\
\lambda^{-1}_{K^+\pi^-}&=&\sqrt{R}\ R_m^{-1}\ e^{-i(\delta+\phi)},\nonumber\\
\lambda_{K^-\pi^+}&=&\sqrt{R}\ R_m\ e^{-i(\delta-\phi)},\nonumber\\
\lambda_{K^+K^-}&=&-R_m\ e^{i\phi}.
\end{eqnarray}
We further define
\begin{eqnarray}\label{defxy}
x^\prime\ &\equiv&\ x\cos\delta+y\sin\delta,\nonumber\\
y^\prime\ &\equiv&\ y\cos\delta-x\sin\delta.
\end{eqnarray}

The processes that are relevant to the CLEO and FOCUS experiments
are $D^0\rightarrow K^+\pi^-$, $D^0\rightarrow K^+K^-$,
$D^0\rightarrow K^-\pi^+$,
and the three CP-conjugate decay processes. We now write down approximate 
expressions for the time-dependent decay rates that are valid for times 
$t\lsim1/\Gamma$. We take into account the experimental information 
that $x$,  $y$ and $\tan\theta_c$ are small, and expand each of the rates only 
to the order that is relevant to the CLEO and FOCUS measurements.
With our assumption that there is no direct CP violation in the processes
that we study, and using the parameterizations (\ref{parlkpi}) 
and (\ref{defxy}), we can write:
\begin{eqnarray}\label{dpikcpv}
\Gamma&[&D^0(t)\rightarrow K^+\pi^-]=e^{-\Gamma t}|A_{K^-\pi^+}|^2
\nonumber\\
 &\times&\left[R+\sqrt{R}R_m(y^\prime\cos\phi-x^\prime\sin\phi)\Gamma t
\right.\nonumber\\
&+&\left.{R_m^2\over4}(y^2+x^2)(\Gamma t)^2\right],\nonumber\\
\Gamma&[&\overline{D^0}(t)\rightarrow K^-\pi^+]= 
e^{-\Gamma t}|A_{K^-\pi^+}|^2\nonumber\\
 &\times&\left[R+\sqrt{R}R_m^{-1}(y^\prime\cos\phi+x^\prime\sin\phi)
 \Gamma t\right.\nonumber\\
&+&\left.{R_m^{-2}\over4}(y^2+x^2)(\Gamma t)^2\right],
\end{eqnarray}
\begin{eqnarray}\label{dkkcpv} 
\Gamma&[&D^0(t)\rightarrow K^+K^-]=e^{-\Gamma t}|A_{K^+K^-}|^2\nonumber\\
&\times&\left[1-R_m(y\cos\phi-x\sin\phi)\Gamma t\right],\nonumber\\
\Gamma&[&\overline{D^0}(t)\rightarrow K^+K^-]=e^{-\Gamma t}|A_{K^+K^-}|^2
\nonumber\\ &\times&\left[1-R_m^{-1}(y\cos\phi+x\sin\phi)\Gamma t\right],
\end{eqnarray}
\begin{eqnarray}\label{dkpicpv}
\Gamma[D^0(t)\rightarrow K^-\pi^+]&=& 
\Gamma[\overline{D^0}(t)\rightarrow K^+\pi^-]\nonumber\\
&=&e^{-\Gamma t}|A_{K^-\pi^+}|^2. 
\end{eqnarray}

\subsection{Experimental results}
The FOCUS experiment \cite{focus} fits the time dependent decay rates of the 
singly Cabibbo suppressed (\ref{dkkcpv}) and the Cabibbo favored 
(\ref{dkpicpv}) modes to pure exponentials. We define $\hat\Gamma$ to be the 
parameter that is extracted in this way. 
The above equations imply the following relations:
\begin{eqnarray}\label{fitexp}
\hat\Gamma(D^0\rightarrow K^+K^-)&=&
 \Gamma\ [1+R_m(y\cos\phi-x\sin\phi)],\nonumber\\
\hat\Gamma(\overline{D^0}\rightarrow K^+K^-)&=&
 \Gamma\ [1+R_m^{-1}(y\cos\phi+x\sin\phi)],\nonumber\\
\hat\Gamma(D^0\rightarrow K^-\pi^+)&=&
 \hat\Gamma(\overline{D^0}\rightarrow K^+\pi^-)\ =\ \Gamma.
\end{eqnarray}
Note that deviations of $\hat\Gamma(D\rightarrow K^+K^-)$ from $\Gamma$ 
do not require that $y\neq0$. They can be accounted for by $x\neq0$ and 
$\sin\phi\neq0$, but then they have a different sign in the $D^0$ and 
$\overline{D^0}$ decays. FOCUS combines the two $D\rightarrow K^+K^-$ modes. 
To understand the consequences of such an analysis, one has to consider the 
relative weight of $D^0$ and $\overline{D^0}$ in the sample. Let us define 
$A_{\rm prod}$ as the production asymmetry of $D^0$ and $\overline{D^0}$:
\begin{equation}\label{defaprod}
A_{\rm prod}\equiv{N(D^0)-N(\overline{D^0})\over N(D^0)+N(\overline{D^0})}.
\end{equation}
Then
\begin{eqnarray}\label{FkkGcpv}
y_{\rm CP}\ &\equiv&\ {\hat\Gamma(D\rightarrow K^+K^-)\over
\hat\Gamma(D^0\rightarrow K^-\pi^+)}-1\nonumber\\
&\approx&y\cos\phi-x\sin\phi\left({A_m\over2}+A_{\rm prod}\right).
\end{eqnarray}
We defined $A_m$ through $R_m^2=1+A_m$ and used the experimental fact
that all $A_i$ are small.
The one sigma range measured by FOCUS is given in eq. (\ref{aveycp}),
\begin{equation}\label{ycpnum}
y_{\rm CP}=(3.42\pm1.57)\times10^{-2}.
\end{equation}

The CLEO measurement \cite{cleo} gives the coefficient of each of the three 
terms ($1$, $\Gamma t$ and $(\Gamma t)^2$) in the doubly-Cabibbo suppressed
decays (\ref{dpikcpv}). Such measurements allow a fit to the parameters $R$, 
$R_m$, $x^\prime\sin\phi$, $y^\prime\cos\phi$, and $x^2+y^2$:
\begin{eqnarray}\label{cleoyx}
R\ &=&\ (0.48\pm0.13)\times10^{-2},\nonumber\\
y^\prime\cos\phi\ &=&\ (-2.5^{+1.4}_{-1.6})\times10^{-2},\nonumber\\
x^\prime\ &=&\ (0.0\pm1.5)\times10^{-2},\nonumber\\
A_m\ &=&\ 0.23^{+0.63}_{-0.80}.
\end{eqnarray}

\subsection{Theoretical Interpretation}
We now assume that the true values of the various mixing parameters
are within the one sigma ranges measured by FOCUS and CLEO. That means
in particular that we hypothesize that $D-\overline{D}$ mixing is being
observed in the FOCUS measurement of $y_{\rm CP}$ and in the CLEO measurement
of $y^\prime\cos\phi$. The combination of these two results is particularly
powerful in its theoretical implications.

Let us first focus on the FOCUS result (\ref{ycpnum}). We argue that it is very 
unlikely that this result is accounted for by the second term in 
(\ref{FkkGcpv}). Even if we take all the relevant parameters to be close to 
their one sigma upper bounds, say, $|x|\sim0.04$, $|\sin\phi|\sim0.6$, 
$|A_m/2|\sim0.4$ and $A_{\rm prod}\sim0.03$, we get $y_{\rm CP}\sim0.01$, about 
a factor of two too small. We can make then the 
following model independent statement: 
\begin{itemize} 
\item if the true values of the mixing parameters are within the one sigma 
ranges of CLEO and FOCUS measurements, then $y$ is of order of a (few) percent. 
\end{itemize}
Note that this is true even in the presence of CP violation, which does allow a 
mass difference, $x\neq0$, to mimic a deviation from the average lifetime. 
Practically, we can take the FOCUS result to be given to
a good approximation by
\begin{equation}\label{newfoc}
y\cos\phi\approx0.034\pm0.016.
\end{equation}
This is a rather surprising result. Most theoretical estimates are well
below the one percent level (for a review, see \cite{hnne}).
These estimates have however been recently criticized \cite{BiUr}.

Second, we examine the consistency of the FOCUS and CLEO results. The two most
significant measurements, that of $y\cos\phi$ in eq. (\ref{newfoc}) and that of
$y^\prime\cos\phi$ in eq. (\ref{cleoyx}) are consistent if
\begin{equation}\label{difocl}
\cos\delta-(x/y)\sin\delta=-0.73\pm0.55.
\end{equation}
This requirement allows us to make a second model independent statement:
\begin{itemize} 
\item if the true values of the mixing parameters are within the one sigma 
ranges of CLEO and FOCUS measurements, then the difference in strong phases 
between the $D^0\rightarrow K^+\pi^-$ and $D^0\rightarrow K^-\pi^+$ decays is 
very large.
\end{itemize} 
For $\delta=0$ we get $y^\prime/y=1$ instead of the range given in eq. 
(\ref{difocl}). To satisfy (\ref{difocl}), we need, for example, 
\begin{equation}\label{delxy}
\cos\delta\lsim\cases{+0.65&$|x|\sim|y|$,\cr -0.18&$|x|\ll|y|$.\cr}
\end{equation}

The result in eq. (\ref{delxy}) is also rather surprising. The strong phase 
$\delta$ vanishes in the $SU(3)$ flavor symmetry limit \cite{woldel}.
None of the models in the literature \cite{ChCh,BrPa,FNP} finds such a large 
$\delta$. Eq. (\ref{delxy}) implies a very large $SU(3)$ breaking effect in 
the strong phase. For comparison, the experimental value of $\sqrt{R}\sim0.07$ 
in eq. (\ref{cleoyx}) is enhanced compared to its $SU(3)$ 
value of $\tan^2\theta_c\sim0.051$ by a factor $\sim1.4$. On the other hand,
there are other known examples of $SU(3)$ breaking effects of order one in
$D$ decays. (For example, $\Gamma(D^0\rightarrow K^+K^-)
/\Gamma(D^0\rightarrow\pi^+\pi^-)=2.75\pm0.15\pm0.16$ experimentally,
while the ratio is predicted to be one in the $SU(3)$ limit.)
So perhaps we should not be prejudiced against a very large $\delta$. 
Furthermore, such a strong violation of $SU(3)$ might help explain
why $y$ is not much smaller than $\sin^2\theta_c$ \cite{bglnp}.

  \section{Conclusions}
Measurements of CP violation provide tests of the Standard Model
and sensitive probes of new physics.  While the information from
near-future experiments will be richer, more accurate and less
subject to hadronic uncertainties, there are several lessons that
can already be drawn from recent measurements:
\begin{itemize}
\item While it is difficult to translate the measured value of
$\varepsilon^\prime/\varepsilon$ into reliable constraints on
the CKM parameters, it is possible to use it in the investigation
of new physics. Some models which predict values that are
substantially smaller than the SM are excluded. In particular,
the superweak scenario is excluded and some models of approximate CP
are disfavored.
\item The measurement of $a_{\psi K_S}$ is not yet accurate enough
to test the Standard Model. It is clear however that the theoretical
cleanliness of this observable will make future measurements very
useful. The situation will be particularly intriguing if the final
result will reside in the lower part of the presently allowed range.
\item The measurements of $D-\overline{D}$ mixing parameters are
not yet accurate enough to test the Standard Model. The values of
these parameters may be below the present experimental sensitivity.
But even if they are within the reach of near future experiments,
present data suggest that they are related to large width difference
and large SU(3) breaking in strong phases rather than to large mass
difference and CP violation. Consequently, there is no hint of new
physics in present data.
\end{itemize}

Hadronic uncertainties play a major role in our various analyses.
We would like to emphasize the following points:
\begin{itemize}
\item It is very important to improve our
knowledge of various hadronic parameters, particularly
$|V_{ub}|$, $\sqrt{B_B}f_B$, $\xi$ and $B_K$.
\item Observables that are very small in the standard model but
likely to be much larger in well-motivated extensions, {\it e.g.}
the electric dipole moment of the neutron, are very useful. 
\item The very few observables that are clean of hadronic
uncertainties, such as the CP asymmetries in $B\to\psi K_S$
and in $K\to\pi\nu\bar\nu$ are crucial in clarifying the picture
of CP violation.
\end{itemize}
 
\acknowledgments

I thank Sven Bergmann, Galit Eyal, Yuval Grossman, Zoltan Ligeti,
Antonio Masiero, Gilad Perez, Alexey Petrov and Luca Silvestrini
for very enjoyable collaborations.


\end{document}